\documentclass[elsart,12pt]{article}
\usepackage{graphicx}
\usepackage{epsfig}
\begin{document}
\baselineskip 20pt plus .1pt  minus .1pt
\pagestyle{plain}
\voffset -2.0cm
\hoffset -1.0cm
\setcounter{page}{01}
\rightline {}
\vskip 1.0cm

\begin{center}
{\Large {\bf PROPERTIES OF COSMIC RAY INTERACTIONS AT PEV ENERGIES}}
\end{center}
\begin{center}
A.D.Erlykin \footnote{Corresponding author: Erlykin A.D., e-mail:
A.D.Erlykin@durham.ac.uk}$^{,2}$, A.W.Wolfendale $^2$
\end{center}
\begin{flushleft}
(1) Department of Physics, University of Durham, South Road, Durham
    DH1 3LE, UK \\
(2) P. N. Lebedev Physical Institute, Leninsky Prosp., Moscow
    117924, Russia \\
\end{flushleft}
 
\begin{abstract}
An analysis has been made of the present situation with the high energy
hadron-nucleus and nucleus-nucleus interaction models. As is already known
there are inconsistencies in the interpretation of 
experimental data on the primary mass composition, which appear when 
different EAS components are used for the analyses, even for the same
experiment. In the absence of obvious experimental defects, there is a
cleari need for an improvement to the exising models; we argue that the most
promising way is to introduce two effects which should be present in
nucleus-nucleus collisions and have not been allowed for before. These are: 
a few percent energy transfer into the EAS electromagnetic component
due to electron-positron pair production or electromagnetic
radiation of quark-gluon plasma and a small slow-down of the
cascading process in its initial stages associated with the extended
lifetime of excited nuclear fragments. The latter process displaces the
shower maximum deeper into the atmosphere.     
\end{abstract}

{\Large {\bf 1. Introduction}}

Among the most popular interaction models which are used for the
analysis of experimental data in the PeV region and beyond are QGSJET,
SYBILL, NEXUS, DPMJET, VENUS and HDPM, coupled with CORSIKA and partly
with the AIRES simulation codes (~see \cite{heck,sciu} and
references therein for a description of the models~).
Early versions of these models differed significantly in their
description of the development of the atmospheric cascades. One of the
possible ways to evaluate the quality of a model is to derive the
primary mass composition from the analysis of different observables:
a good model should obviously give the same composition for any set of
observables.
We used this approach to interpret the data on
the maximum development depth of the electromagnetic cascades $X_{max}$ and on
the ratio of muon to electron sizes $N_{\mu}/N_e$ in terms of the
mean logarithm of the primary mass $\langle lnA \rangle$; the best
consistency was obtained for the QGSJET model \cite{EW98}. A later analysis of
the EAS hadron component confirmed this conclusion \cite{anto99}. 

Since then the models have been improved and the differences between their
predictions reduced \cite{heck}. The precision of
experimental data has been also improved and this offers the hope of
further testing the models and of determining the actual value of
$\langle lnA \rangle$. Some of the new measurements in which
Cherenkov light in the atmosphere is detected
have given a deeper $X_{max}$ and hence a lighter mass composition
than before \cite{swor,fowl}. The
extensive set of measurements on the ground and relevant simulations made
by the KASCADE collaboration indicated that there is a difference in the
estimate of $\langle lnA \rangle$ derived from different sets of observables 
\cite{roth}. At lower energies it was also difficult to
get a consistent explanation of the measured muon and hadron trigger 
rates \cite{anto01}. All these facts provide evidence that
further improvements to the models are urgently needed.  

Meanwhile, a few calls for a quite radical change of the interaction
model have appeared, inspired mostly by the existence of the well known knee in
the primary cosmic ray (~CR~) spectrum at PeV energies. We discuss
first the need for such a radical
change and follow this with our own suggestion for modifications to the
model for a better description of the modern experimental data 
(~{\em i.e.} KASCADE, DICE, BLANCA {\em et al.}~). 

{\Large  {\bf 2. The possibility of a radical change in the
interaction model}}

{\large {\bf 2.1. The possibilities and general remarks}}

A number of possibilities arise, and they will be considered in turn.
Essentially they are of two types. The first considers that the
primary spectrum (~represented in the usual way by a power law~) has
an energy-independent exponent valid both below
and above the PeV region where there is the knee in the spectrum
inferred from the detected
particles 'low' in the atmosphere (~electrons, muons, hadrons~). The
change in the interaction mechanism is then sudden (~at $\approx$ 3
PeV~) for CR-air nucleus interactions. The second is of the same
philosophy i.e. that the original production spectrum has a unique
exponent, but the interactions are in the interstellar medium (~ISM~)
or even in the intergalactic medium (~IGM~)
and not with air nuclei so that the spectrum of CR incident
on the atmosphere already has the knee in it; there is no further
radical change in the interaction mechanism for CR - air nucleus
collisions.
  
The general argument against all these ideas is based on the observed    
anisotropy of the arrival directions of cosmic rays. The measured amplitude and
phase of the anisotropy change in the vicinity of the knee
and this gives support for its astrophysical origin \cite{clay,ELW98,EW01},
i.e. that the spectral shape arises because
several sources, or types of source, contribute.
 
Of relevance, too, is the cosmological aspect, {\em viz} the effect of
a proposed radical interaction change in the early Universe when the average
energy of all 'particles' was in the PeV region. As we have pointed
out already \cite{EW98} there are sound cosmological arguments
against a radical change for P-P reactions in the PeV region. Some
10$^{-14}$ sec after the Big Bang the average temperature of the
primordial plasma was about 10$^{16}$K and the mean energy per
'particle' was of the order of a TeV in the centre of mass system, which 
corresponds to PeV energies in the laboratory system. If a radical
change of the interaction occured at this energy changes to the subsequent
evolution of the Universe would have taken place \cite{bere}. Now the contemporary
model (~i.e. without a radical change~) fits the observed data on the
relative abundance of the light elements and it is difficult to
imagine this being the case were there to have been a radical change
of the interaction at 10$^{-14}$ s. 

Despite 'the case against' the possibility of such a radical change it
would have such
important implications for particle physics that we make a detailed
examination of some of the
recent radical interaction models put forward to explain the knee.

{\large {\bf 2.2. A radical change for the CR - air nucleus
interaction}}

Nikolsky has argued extensively that there is such a radical change.
The latest idea \cite{niko01} is close to that of his paper
of 1995 \cite{niko95}. The author
assumes that at a primary energy of about 3-5 PeV the multiplicity of secondary
particles increases dramatically. The attenuation of the cascade
speeds up and the intensity of showers in the lower half of the atmosphere
falls, thereby forming the knee. There is actually no detailed model to
be checked, because no numerical estimates have been
given of the basic interaction parameters and their energy dependence, 
therefore it is difficult to assess the idea quantitatively.

One remark should be made, however. There is a difference between the
previous and present pictures of the interaction advocated by the
author. The 1995 model had the need to withdraw a part of the
cascade energy and to put it into some unobservable components (~high
energy muons, neutrinos etc.~) in order to form the knee. Now, in 2001,
on the other hand, more energy is transferred to low energy particles
of the hadronic cascade in the initial stages of its development. In
our mind this is a backward step.
The suggested increase of the energy transferred into the hadronic cascade in
the atmosphere should be seen in the spectra of muons and Cherenkov light 
emitted by the charged particles in 
the atmosphere as an 'inverse knee or ankle', i.e. as a {\em
decrease} of the spectral slope. Nothing like that is
actually observed. EAS muon size spectra at mountains: Tien-Shan 
\cite{stam,roma}, EAS-TOP \cite{nava,agli} and at sea level: Ohya
\cite{mits}, KASCADE 
\cite{glas,haun} and  MSU \cite{fomi} all show the 'ordinary' knee,
i.e. an {\em increase} of the spectral slope. 
To the existing 5 Cherenkov light spectra: TUNKA \cite{gres}, HEGRA 
\cite{arqu}, DICE \cite{swor}, BLANCA \cite{fowl} and CACTI
\cite{pali} has been added another one which is based on the
combined use of Cherenkov light and EAS electron size measurement 
\cite{knur}. All 6 demonstrate clearly the existence
of the knee just like that in the particle size spectrum with a
sharp {\em increase} of slope, which rules out the possibility of
radical increasing the energy fraction deposited by the cascade in the 
atmosphere.

The possibility of having no knee in the primary spectrum and hiding
the missing energy in high energy muons and neutrinos ( an apparently 
attractive idea ) has also been discussed by Petrukhin \cite{petr}. The
proposed model
suggests the production at threshold of one or a few hypothetical heavy
particles in each interaction, which then decay into muons and
neutrinos. The energy flux contained in his muon spectra 
is less than the missing energy flux, needed to form the knee, by a
factor of about 20 even for the case of the maximum flux of high
energy muons. This means that the bulk of the missing energy
should be carried away by neutrinos. The author succeeds in making the energy
spectrum of muons from the hypothetical decay much harder than that of 
the background muons from $\pi$,$K$ decays and prompt muons. In this way
it is possible to make the missing energy
visible only as a very low flux of extremely high energy muons.
However, such a model of the 'new physics' ignores everything we 
know about the multiple production at high energies ( small
cross-section for the production of heavy particles, conservation of
lepton number in weak decays, etc.) and seems hardly credible.

Another model of 'new physics' discussed by Kazanas and Nicolaidis 
\cite{kaza}, 
in spite of some theoretical arguments, is actually purely
phenomenological. In it one opens a new channel for
multiple production and puts part of the primary energy into some
unobserved particles ( technihadrons, gravitons  ), then, if one fits the
energy dependence of the cross-section for the new channel and the 
energy fraction contained in the new particles, one can get a reasonable
fit to the observed primary cosmic ray energy spectrum, including the 
knee. It is difficult to argue against the fit, but there seems to us
to be no support for such speculation.

{\large {\bf 2.3. Radical changes in the interactions in the
interstellar medium}}
    
It is, in principle, possible that the knee is due to interactions of
some form beyond the atmosphere - in the ISM or IGM. 
If, as might be assumed by analogy with the 'X-particles' produced by
cosmic strings, the electromagnetic component is a prominent
end-product of this interaction, then this hypothesis can be ruled out 
because the ensuing
cascade in the Universe (~inverse compton scattering on the cosmic
microwave background~) would produce a gamma ray intensity in excess
of that measured (~see, {\em e.g.} \cite{chi} for the related
cosmic string arguments~). However, there are other possibilities for 
the radical change of the interaction, which can now be considered. 

The possibility of finite neutrino mass has given birth to models
in which the knee is caused by the interaction of the primary protons
with massive relic neutrinos. 
Assuming that the mass of the electron antineutrino is 0.4 eV/c$^2$,
Wigmans \cite{wigm} converts some of the primary protons by inverse
$\beta$-decay reaction into neutrons and positrons, starting from an
energy of about 4 PeV. Indeed, protons will disappear from the primary
flux, but we ourselves argue that even if the shower from a secondary 
positron misses the detector, the shower
produced by the neutron will be practically indistinguishable from a
proton-initiated shower, because neutrons preserve more than 99\% of the
primary proton energy; thus EAS measurements would not see a
knee. Taking into account the problem of the very small 
cross-section for the neutrino induced reactions, which requires us to
assume a very dense halo of neutrinos gravitationally attracted to
the cosmic ray source region (~be it the Galaxy or source within the
Galaxy~), one should remark that the PeV neutron will decay again into
a proton and a positron within tens of parsecs and the model loses its
value.            
       
Another model of the intermediate type was proposed by Ehrlich \cite{ehrl},
He used some theoretical ideas that the electron neutrino is a
tachyon, which in fact does not contradict the existence of a finite
neutrino 
mass. 
In this case, starting from the same energy of 4 PeV, the primary
proton has the possibility to decay into a neutron, a positron and
an electron neutrino. This decay might create a narrow peak of neutrons
in the primary cosmic ray flux in the region close to 4.5 PeV, which
can perhaps imitate the knee. All the scepticism put forward
in the previous paragraph is applicable also in this case, however. 
  
The common deficiency of all the models on the radical change of the 
interaction is that no
attempts have been made to reconcile it with the multitude of other 
experimental data, different from those used for the motivation of the
model. The conclusion which can be drawn from this survey is that 
at the present time {\em there is
neither a reasonable model nor solid argument which justifies  
a \underline {radical} change of the interaction.} We conclude that
such a change as is necessary for the interaction model should not be
radical and not responsible for the origin of the knee, but only
responsible for restoring consistency for all cosmic ray measurables.

{\Large {\bf 3. What kind of the modification of the interaction model
do we need ? }} 

The variety of observed EAS components and precise measurements of their
characteristics by the KASCADE experiment allowed this collaboration to
derive the primary mass composition using the multivariate analysis of
data. Remarkably, as has been pointed out already, it was
found that not only do different methods applied to the same set of 
observables give different results, but that there is a systematic
difference between the results obtained using the same method, but
different components.  
In Figure 1 the mean logarithm of primary mass $\langle lnA \rangle$
derived from different sets of observables is shown vs. the so called {\em
truncated} muon number $N_\mu^{tr}$ (~the figure is taken from
\cite{roth}~). $N_\mu^{tr}$ is the number of low energy muons 
(~$> 0.25 GeV$~) contained
between 40 and 200 m from the shower axis and is an observable adopted by the
KASCADE collaboration as a measure of the primary energy, independent
of the primary mass. They also use the number of higher energy muons 
(~$> 2.4 GeV$~) $N_\mu^\ast$ as well as some observables related to
the hadron component for the analysis of the mass composition
\cite{anto99}. The fundamental problem is, of course, that all the
$\langle lnA \rangle$ values should be the same, at the same value of
$N_\mu^{tr}$, and they are not.

\begin{figure}[htbp]
\vspace{0.5cm}
\begin{center}
\includegraphics[height=18cm,width=15cm,angle=0]{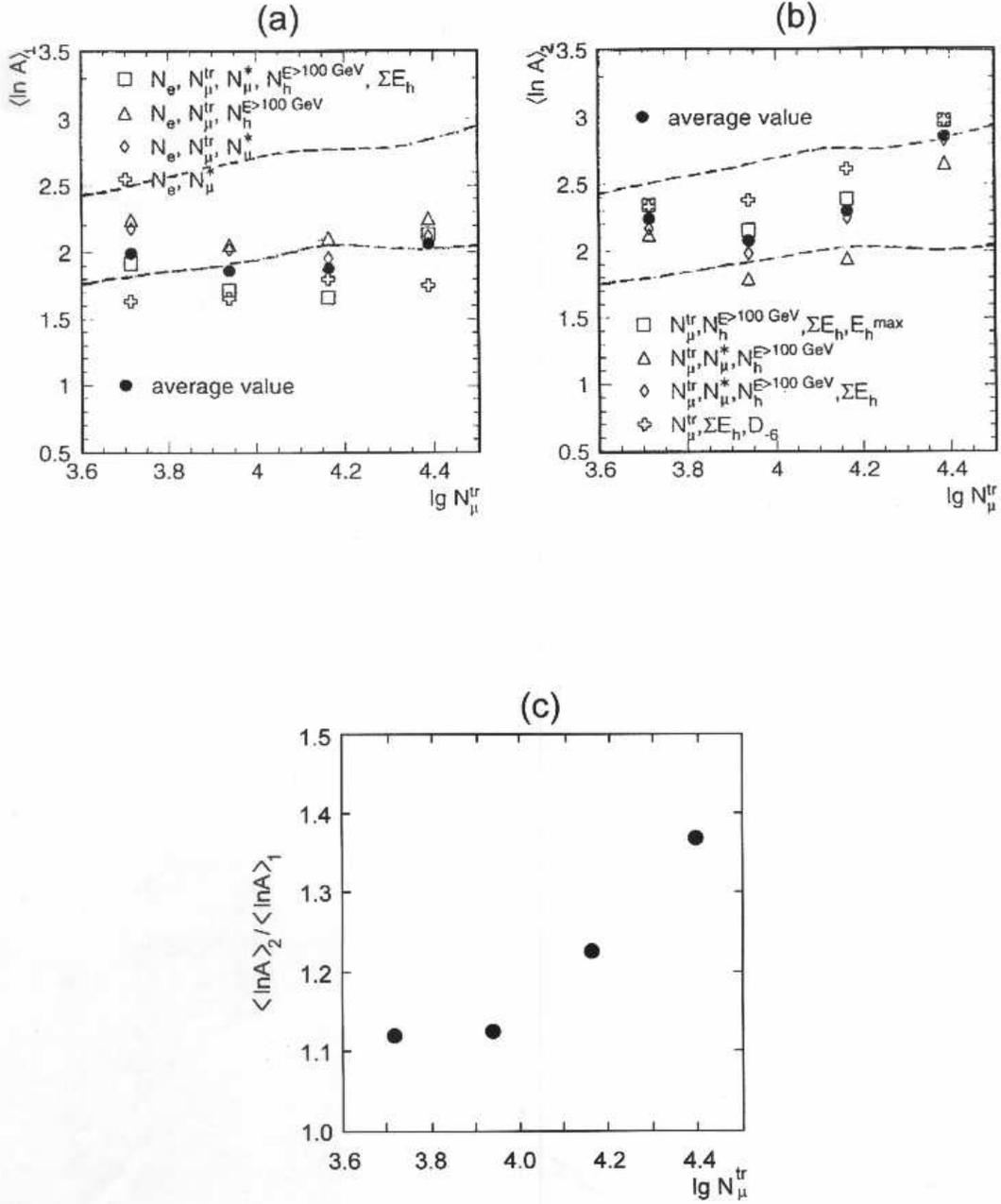}
\caption{ \footnotesize (~a,b, taken from \cite{roth}~). 
Mean logarithmic mass
$\langle lnA \rangle$ from the analysis of different sets of
observables vs. $lgN_\mu^{tr}$ ( QGSJET prediction ). The 
two dashed lines indicate alternative predictions of the Single Source Model 
\cite{EW98}, with the upper line being our latest and 'best
estimate'. The sets displayed in (b) do not include the observable
$N_e$. It is seen that omitting $N_e$ results in a heavier mass composition.
(c) The ratio of mean values of $\langle lnA \rangle$ from sets (b)
and (a).
We argue, in the text, that the 'true' value of 
$\langle lnA \rangle$ is a little below the crosses in Figure 1b. }
\end{center}
\label{fig:int1}
\end{figure}

A number of features are of interest and, presumably, of some
importance.

(i) The dispersion of the individual $\langle lnA \rangle$ values
about smooth lines through them indicates that non-systematic errors are
$\approx \pm$0.05. 

(ii) The use of the electron size $N_e$ results in an average lighter 
composition (~Fig.1a~). On the other hand, omitting electrons and using
just muons and hadrons results in a heavier composition (~Fig.1b~). 

(iii) Considering the mean values from the two sets of data (~filled-in
circles~), the ratio of $\langle \langle lnA \rangle \rangle$ for (b)
to (a) increases smoothly with $lgN_\mu^{tr}$, i.e. the discrepancy
between (b) and (a) rises with increasing primary energy (~Fig.1c~). 

(iv) Within each of (a) and (b) the differences between the $\langle
lnA \rangle$ values using different combinations of data (~$N_e,
N_\mu^\ast$ etc.~) are systematic and real (~see (i)~).

(v) In both (a) and (b), the highest estimated $\langle lnA
\rangle$ values are for the cases where $N_\mu^\ast$ is absent - 
i.e. not used in the analysis. The lowest values of $\langle lnA
\rangle$ are where 
$N_\mu^\ast$ has the greatest weight in the analysis, i.e. the number
of parameters (~2 in 1a and 3 in 1b~) is a minimum. It is an
indication that the higher energy muons (~$>2.4 GeV$~) in the observed
showers are in the greater deficit compared with model predictions
\cite{haun}. 

(vi) With respect to (v), taking only the two sets (~one in 1a and one
in 1b~) which did not use $N_\mu^\ast$, it is interesting to note that
the former points are close to our lower variant (~lower dashed
curve~) in 1a and close to our 'best estimate' (~upper dashed curve~) in
1b. The differences are, therefore, significant.

The difference in the $\langle lnA \rangle$ values derived using
different observables (~Fig.1~) points to inadequacies in the models 
used for the analysis of the experimental data and provides an impetus 
to correct them. However, as remarked already, it
is sure that the corrections should not be radical, because the
difference between the $\langle lnA \rangle$ is not large, typically 
$\delta \langle lnA \rangle \approx 0.2 - 0.4$ or, in terms of the
ratio of the mean
$\langle lnA \rangle$ values from Figures 1b and 1a, an increase
from 1.12 to 1.37. Below, we show that the mentioned systematic 
difference between the $\langle lnA \rangle$ values can
be an indication that the energy distribution between the different shower
components is slightly different from that in the models: specifically,
the actual mean number of electrons $N_e$ in EAS appears to be higher, that
of muons $N_{\mu}$ slightly
lower and that of hadrons $N_h$ lower still than in the models. 

It is well known from the present models that
for the same primary energy the number of muons in nuclei-induced
showers is higher than in proton-induced ones. On the other hand the
number of electrons and hadrons in nuclei-induced showers observed in
the lower half of the atmosphere is lower than in proton ones
\cite{knap}. If one has an opportunity to measure the primary
energy, by muons, Cherenkov light or another technique, and finds that
the shower has a low $N_{\mu}$ or a high $N_e$ (~actually the ratio
$\frac{N_{\mu}}{N_e}$ is important~) the conclusion will be that
this shower is initiated by a proton or light nucleus. On the
contrary, if one
finds that the shower has a low $N_h$ the conclusion will be the 
opposite, i.e. that the shower is initiated by a heavy nucleus. This is
exactly what is observed in the showers at sea level. 
In order to get a better consistency between the 
$\langle lnA \rangle$ values obtained using $N_e$, $N_\mu^{tr}$ 
and $N_h$ ( Fig.1 ) 
the model should have higher $N_e/N_{\mu}^{tr}$ and smaller           
$N_h/N_{\mu}^{tr}$ ratios. This situation could be achieved if the corrections
give a reduction of $N_h$, a smaller reduction of $N_{\mu}^{tr}$ and  
an increase in $N_e$.

The same conclusion on the improvement of the model can be drawn from
an analysis of KASCADE event rates \cite{anto01}. Both the muon
and hadron trigger rates, observed by KASCADE, are lower than expected
from the model calculations. This discrepancy indicates again that the
actual 
numbers of muons and hadrons in EAS are lower than in the models, although the
energy region responsible for the trigger rates is lower than that
analysed for the mass composition around the knee. The argument that
the lower trigger rates might be due to a lower primary intensity is 
disproved by the fact that the
ratio of hadron to muon trigger rates, which is independent of the absolute
intensity, is also lower than in the calculations \cite{riss}.  
To get agreement the predicted muon trigger rate should be reduced
by a factor 0.89 $\pm$ 0.06, and the hadron rate by 0.54 $\pm$ 0.08
compared with the rates given by the QGSJET model. This
analysis is of value because it indicates that the needed
reduction of the number of muons in the model should be about 6\%, for
hadrons it is bigger - about 29\%. Because muons and hadrons are the
products of hadronic cascades, it is evident from the energy
balance that to reduce the energy contained in the hadronic cascade
one has to increase the energy transferred to the electromagnetic cascade.  
At this stage it can be remarked that an increase in this energy is
predicted by theory ( although it appears not to have been included
into the models hitherto ).
\newpage
{\Large {\bf 4. Numerical estimates}}

{\large {\bf 4.1. The balance between the EAS components}}

To evaluate the effect of the proposed change of the balance between 
different cascade
components we applied the semi-quantitative analytical approach. We
assumed that there are only nucleons (N), pions ($\pi$),
muons+neutrinos ($\mu \nu$) 
and the electromagnetic component ($e\gamma$) in the cascade. The total
and partial inelasticity for nucleons: $K^N_{tot}$ and $K^N_{\gamma}$
and for pions $K^{\pi}_{tot} = 1$ and $K^{\pi}_{\gamma}$, interaction
mean free paths for nucleons $\lambda_N$ and for pions $\lambda_{\pi}$
 were taken to be energy independent, {\em i.e.} constant. The system
of kinetic
equations, describing the longitudinal development of the nucleon,
pion, muon+neutrino and the electromagnetic energy of the cascade has
been taken as \\
$\frac{dE_N}{dX}=-\frac{E_N}{\Lambda_N}$ \\ 
$\frac{dE_{\pi}}{dX}=-(\frac{1}{\Lambda_{\pi}}+\frac{E_{cr}}{\langle
E_{\pi} \rangle
X})E_{\pi}+\frac{K^N_{tot}-K^N_{\gamma}}{\lambda_N}E_N$ \\ 
$\frac{dE_{\mu \nu}}{dX}=\frac{E_{cr}}{\langle E_{\pi} \rangle X}E_{\pi}$ \\
$\frac{dE_{e\gamma}}{dX}=\frac{E_{\pi}}{\Lambda_{\pi}}+\frac{K^N_{\gamma}}{\lambda_N}E_N$
\\
Here X is the atmospheric depth, $\Lambda_N = \frac{\lambda_N}{K^N_{tot}}$,
$\Lambda_{\pi}=\frac{\lambda_{\pi}}{K^{\pi}_{\gamma}}$, $E_{cr}$ - the
critical energy for pion decay in the air, $\langle E_{\pi} \rangle$ -
the mean energy for pions. The boundary condition at $X=0$ is  
$E_{\pi}+E_{\mu \nu}+E_{e\gamma}=0$ \\
$E_N=E_0$ \\
where $E_0$ is the total energy of the cascade. The solution of the
system of equations is \\
$E_N=E_0exp(-\frac{X}{\Lambda_N})$ \\
$E_{\pi}=E_0
\frac{K^N_{tot}-K^N_{\gamma}}{\lambda_N}exp(-\frac{X}{\Lambda_{\pi}}) \int_0^X(\frac{y}{X})^{\frac{E_{cr}}{\langle
E_{\pi} \rangle}} exp(-\frac{y}{\Lambda_{\pi N}}) dy$ \\ 
$E_{\mu \nu}=\int_0^X \frac{E_{cr}E_{\pi}(y)}{\langle E_{\pi} \rangle
y} dy$ \\
$E_{e\gamma}=\int_0^X \frac{E_{\pi}(y)}{\Lambda_{\pi}}dy +
\frac{E_0K^N_{\gamma}}{K^N_{tot}}(1-exp(-\frac{X}{\Lambda_N}))$ \\
Here $\Lambda_{\pi N}=\frac{\Lambda_{\pi} \Lambda_N}{\Lambda_{\pi}-\Lambda_N}$.

We calculated also the longitudinal development of the electron size
of the shower $N_e$ as \\
$N_e(X)=\int_0^X \frac{dE_{e\gamma}}{dy}\tilde{N}_e(X-y)dy$ \\
The longitudinal profile of the electromagnetic cascade has been taken
as \\
$\tilde{N}_e(t)=\frac{E_{e\gamma}}{1GeV}exp(t-t_{max}-2tlnS)$,
where the atmospheric depth $y$ is measured in units of the radiation
length $t_0$: $t = \frac{y}{t_0}$,
$t_{max}=1.7+0.76ln(\frac{E_{e\gamma}}{\beta})$,
$S=\frac{2t}{t+t_{max}}$ \cite{cata}.
Here $E_{e\gamma}$ is the total energy of the electromagnetic cascade, $t_0$
= 37.1 gcm$^{-2}$ and $\beta$ is the critical energy for the
electromagnetic cascade in air, equal to 0.081 GeV. Although this
approximation is recommended for nuclear cascades, its use for the
electromagnetic cascades is more appropriate for our energy-balance
approach, since it gives the correct estimate of the depth of the
shower maximum (~$X_{max} \approx 610 gcm^{-2}$~) and its elongation
rate (~$ER \approx 65 gcm^{-2}$~).

The integrals for $E_{\pi}$, $E_{\mu \nu}$ and $E_{e\gamma}$ were
calculated numerically. The basic parameters for the calculation were
taken as
$K^N_{tot}$=0.60, $K^N_{\gamma}$=0.20, $K^{\pi}_{\gamma}$=0.33,
$\lambda_N$=90 gcm$^{-2}$, $\lambda_{\pi}$=120 gcm$^{-2}$, $E_{cr}$ =
90 GeV, $\langle E_{\pi} \rangle$ = 300 GeV.

The results of the calculation are shown in Figure 2 by the full line.
According to our suggestion, we increased the energy fraction
transferred by nucleons into the electromagnetic component, leaving
all the other characteristics untouched. We increased $K^N_{\gamma}$
from 0.20 to 0.26 ( the reason for this particular value will be clear
later ). The result is shown in Figure 2 by the dashed line:
 the muon energy at sea level (~1000 gcm$^{-2}$~) decreased by
$\sim$6\%, the hadron energy decreased by $\sim$23\%, the energy transferred
into the electromagnetic component increased by $\sim$2\%.

\begin{figure}[htbp]
\vspace{0.5cm}
\begin{center}
\includegraphics[height=18cm,width=15cm,angle=0]{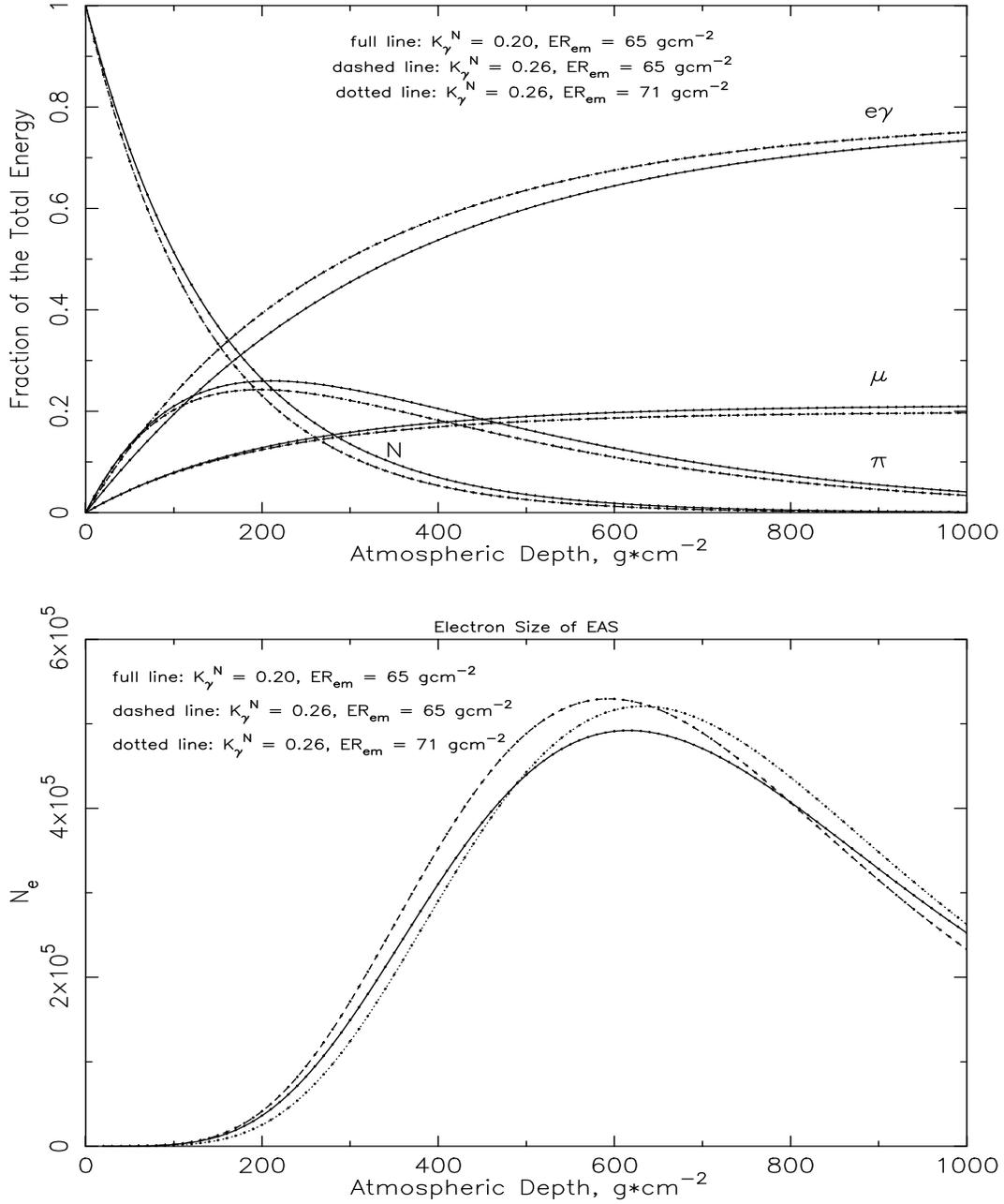}
\caption{\footnotesize The longitudinal development of 1 PeV cascades in the 
atmosphere: (a) Fractions of the total energy carried by nucleons
($N$), pions ($\pi$), muons ($\mu$) and transferred into the
electromagnetic ($e\gamma$) component ; (b) Electron Size of the
shower $N_e$. Basic parameters of the calculation are given in the
text. Full line: $K^N_{\gamma} = 0.20, ER = 65 gcm^{-2}$; 
dashed line: $K^N_{\gamma} = 0.26, ER = 65 gcm^{-2}$;
dotted line: $K^N_{\gamma} = 0.26, ER = 71 gcm^{-2}$. }
\end{center}
\label{fig:int2}
\end{figure}

{\large {\bf 4.2. The triangle diagrams}}

The balance of the energy contained in the major components of the
shower is convenient to analyse using the so-called ``triangle diagrams'' 
\cite{dani83,dani93}. If the height of
an equilateral triangle is equal to 1, then for each point inside this
triangle the sum of the distances to its sides is equal to 1. If we
know the energy fractions carried by the electromagnetic ($\delta_{e\gamma}$),
muon ($\delta_{\mu}$) and hadron ($\delta_h$) components of the shower
at the observation level, so that
$\delta_{e\gamma}+\delta_{\mu}+\delta_{h}=1$, then each shower can be
presented by a single point inside the triangle. Our basic shower 
( Figure 2, full line ) is shown in Figure 3 by a full circle
(~$\delta_{e\gamma}$=0.394, $\delta_{\mu}$=0.505, $\delta_h$=0.101~). The
desired direction for the shift of the energy balance in the
modified model is shown by the straight arrow in Figure 3b.    

\begin{figure}[htbp]
\vspace{0.5cm}
\begin{center}
\includegraphics[height=18cm,width=15cm,angle=0]{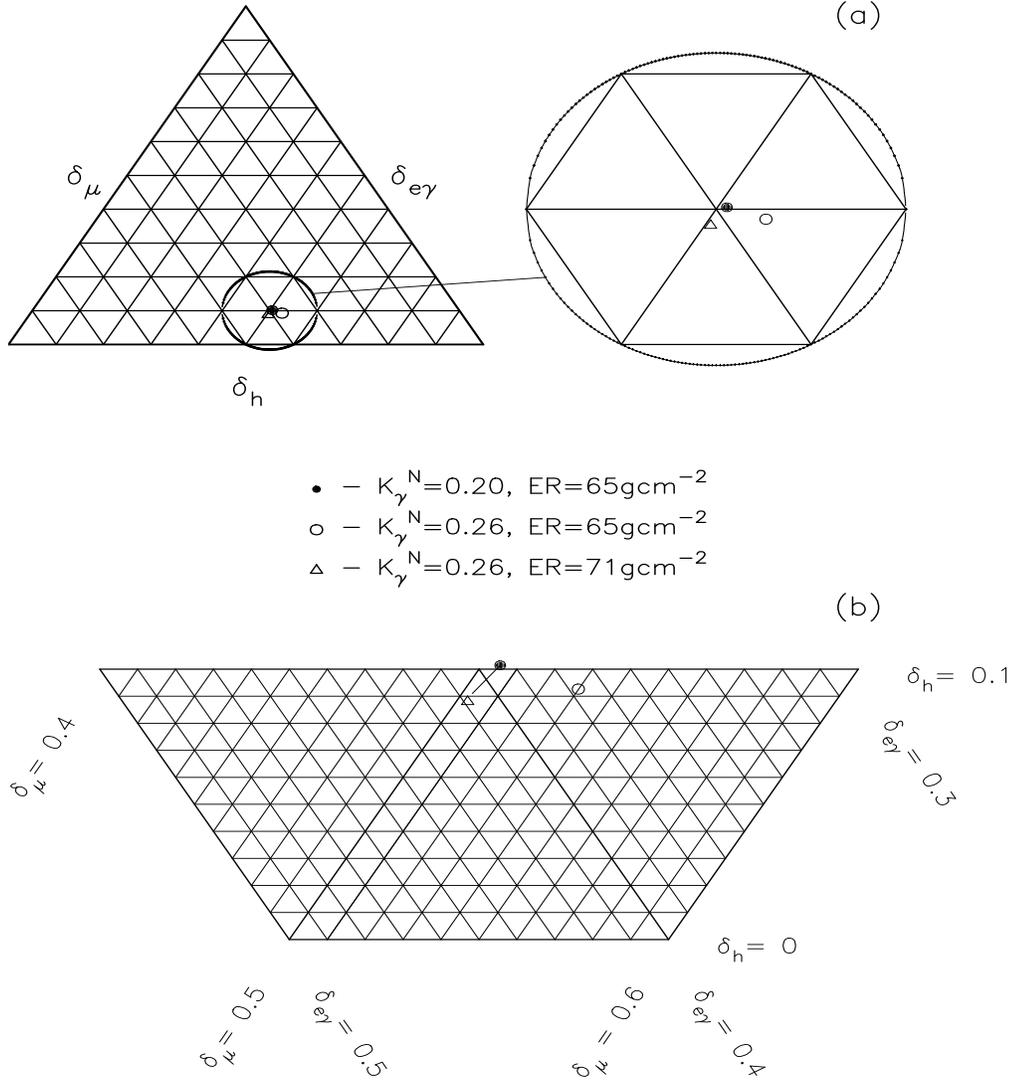}
\caption{\footnotesize  Triangle diagram for a 1 PeV shower at the sea
level: $\delta_h, \delta_{\mu}, \delta_{e\gamma}$ are fractions of
energy carried by the hadron, muon and electromagnetic components of the
shower, respectively. $\bullet$ - $K^N_{\gamma} = 0.20,
K^N_{tot} = 0.60, ER = 65 gcm^{-2}$; $\bigcirc$ - $K^N_{\gamma} =
0.26, K^N_{tot} = 0.66, ER = 65 gcm^{-2}$; $\triangle$ -
$K^N_{\gamma} = 0.26, K^N_{tot} = 0.66, ER = 71 gcm^{-2}$.
(a) Large scale diagram with the inset magnifying the indicated region.
(b) Small scale part of the full diagram. The straight arrow to the left of the
full circle indicates the desired modification of the EAS energy balance. }
\end{center}
\label{fig:int3}
\end{figure}

However, despite the increase of the energy {\em transferred} into the   
electromagnetic component, the {\em preserved} electromagnetic energy
and the electron size of the shower $N_e$ at sea level decreased by
14\% due to its faster development and
then faster attenuation of the cascade. The point in the triangle
diagram moved in the different direction (~open circle~). The shift is similar to the case when we
increase the total inelasticity $K^N_{tot}$ and $K^N_\gamma$ with
$K^N_\gamma = \frac{1}{3}K^N_{tot}$. 

If we were to assume a larger fluctuation $\sigma(K^N_{\gamma})$ compared with
those in the model this would result in a wider distributions of $N_e$, 
$N_{\mu}$ and $N_h$
and in a bigger value of $\sigma(lnA)$. Within our approach we cannot give the
numerical estimates, but we expect that the mean energy balance would
not move in the desired direction and the mean value $\langle lnA
\rangle$ would not change significantly. Therefore, {\em an increase 
of the mean value or increased fluctuations of $K^N_{\gamma}$ cannot give the 
required result.}

{\large {\bf 4.3. Data from mountain laboratories}}

We should remark that the conclusion about the decrease of $N_e$
relates only to measurements at sea level. At mountain altitudes the 
increase of $K^N_{\gamma}$ results in an increase in $N_e$ as seen in Figure
2, however it is not enough here, either. The best complex EAS array at
mountain altitude
was at Tien-Shan and the
data there also indicated the need for the model to be improved 
\cite{erly94}. The problem was with the so-called 'observed' mass 
composition, i.e. the mass composition of primary particles responsible
for showers at the observation level selected by one of their
parameters. The Tien-Shan array is 
at a depth of 690 gcm$^{-2}$, i.e. PeV-showers are definitely
observed beyond their maximum. The $N_e$-selected showers have
to be enriched by showers initiated by protons and light nuclei,
while in contrast the $N_{\mu}$-selected showers have to be
enriched by heavy nuclei initiated showers. However, in fact, the 'observed'
composition at Tien-Shan was about the same for these two selections 
\cite{stam,erly94}. The consistency can be restored by
the same way as in KASCADE, i.e. by changing the balance between the muon and
electromagnetic components in the models, in favor of the latter. 

{\large {\bf 4.4. Slowing down the cascade}}

We have concluded that the mere increase of the energy fraction 
$K^N_{\gamma}$ transferred into the electromagnetic component does not
change the balance properly, although it does help in some respects
and, as we have stated, it {\em must} be present.  
We now argue that another effect that {\em should} be present will
bring about the desired effect: the slowing down of the development of
the cascade in its initial stages. The physics behind this
modification will be discussed later. Here we just demonstrate
the validity of this argument in a semi-qualitative way. For
illustration purposes we slow down the development of the hadronic and 
electromagnetic
cascade by increasing the elongation rate 
$ER$ from 65 gcm$^{-2}$ to 71 gcm$^{-2}$, preserving
$K^N_{\gamma}$ = 0.26. The result is
shown in Figure 2 by dotted lines and in Figure 3 by an open triangle.
The direction of the shift is correct and its magnitude can be
adjusted to achieve consistency in the mass composition.

Thus we conclude that {\em the increase of the energy transferred into the
electromagnetic component (~$K^N_{\gamma}$~) combined with the
slowing down of the development of cascades in their initial stages
is the most realistic way to improve the particle interaction model
and to achieve a consistent estimate of the primary mass composition.}    

{\Large {\bf 5. Theoretical arguments}}

{\large {\bf 5.1. General remarks}}

All the arguments hitherto were purely phenomenological. However,
there are theoretical arguments too which lend support to the
phenomenological consideration as will be demonstrated. At the moment 
nearly all the models
for hadron-nucleus ({\em hA}) interactions implemented in CORSIKA are well
developed and theoretically justified. Nucleus-nucleus ({\em AA})
interactions are considered within the framework of the Glauber approach,
which reduces the $AA$-collision to the sum of a few $hA$-collisions.
No new processes are assumed to occur in $AA$-collisions. 
However, there {\em are} processes which must be present but which have not
been included hitherto, and we think that both effects: the increase
of $K_{\gamma}$ and the slowing down of the cascade development 
refer to processes, intrinsic for $AA$-collisions, which will cause at least
near-equality of the $\langle lnA \rangle$ values in Figures 1a and 1b.

{\large {\bf 5.2. The increase of $K_{\gamma}$}}

At PeV energies, when the
Lorenz-factor of the projectile nucleus approaches $\sim 10^6$
and its charge $Z$ is $\sim 10-30$, the density of virtual photons
in the contracted coulomb field is so high that when the projectile collides
with a nitrogen ( $Z$ = 7 ) nucleus of the air and the 
impact parameter is small (~central collision~) the probability of 
electron-positron pair production including multiple pair production 
becomes quite appreciable \cite{erly75,baur,lee}. Using the numerical estimates
made by Erlykin \cite{erly75} for p-air collisions, we can extrapolate them
to Fe-air collisions and expect an increase of
$K^{Fe-air}_{\gamma}$ for the average collision at PeV energies by
0.05-0.13. For specific central collisions the effect will
be even higher. Our 0.06 of \S4.1 was thus rather conservative.
We also emphasize the fact that ordinary $K^{AA}_\gamma$ value in
AA-collisions is not very high. Typically for Fe-air collisions the
average number of wounded nucleons $n_w$ is about 7-8 and 
$K^{Fe-air}_\gamma \approx K^{PA}_\gamma \frac{n_w}{A_{Fe}} \approx
0.03$. Therefore even a few percent increase of
$K^{AA}_\gamma$ in the first interaction can be essential for the
subsequent development of the atmospheric cascade. 

Experimental results confirm the intense production of
electron-positron pairs in $AA$ collisions. In the CERES/NA45 experiment
the number of $e^+e^-$ pairs in the low invariant mass region
$m_{e^+e^-} < 200 MeV/c^2$ exceeds the number in the higher mass
region of $m_{e^+e^-} > 200 MeV/c^2$ by a factor of $\sim 10^4$. For
the CERES/NA45 collaboration it is just a background, because they are 
interested in dileptons with high invariant mass \cite{agak,fili} 
as a signature of an elusive quark-gluon
plasma.  We believe that {\em to check our assumption
the study of
the dilepton production should be concentrated preferentially in the
region of invariant mass below} $10 MeV/c^2$.
 
An additional energy transfer into an electromagnetic component can
arise also from an excess of direct photons, which has been predicted
theoretically as a signature of the quark-gluon plasma \cite{fein} 
and is now observed in AA-collisions both at low and high 
transverse momenta \cite{agga,peit}.

Another theoretical possibility is to increase the transfer of the energy
into the electromagnetic component, by way of  
$hA$-interactions; this could be connected with the breaking of the diquark and
its subsequent recombination into
3 $\pi^0$-mesons \cite{cape,capd}.
     
{\large {\bf 5.3. Slowing down of the cascade development}}

The theoretical motivation of the second assumption is also connected
with the properties of AA-interactions, but relates mainly to
peripheral collisions. It is known that even at high energies a
projectile nucleus does not fully disintegrate into constituent
nucleons in the first interaction but fragment into few pieces of
different mass \cite{gior}. Some of these fragments are excited and,
after the de-excitation, if it occurs in space, give rise to MeV gamma-ray
lines, observed from 'discrete' sources and the interstellar medium 
\cite{rama}. The lifetime of the 
excited fragments varies from a 'nuclear' time $\sim 10^{-23}$ sec 
to millions of years. For AA-interaction at PeV energies both the lifetime
before de-excitation and the energy of emitted gamma-quanta are
extended by the factor of $\sim 10^5 - 10^6$ due to relativistic effects. 
As a consequence one can expect an additional sub-PeV electromagnetic cascade
to be initiated a few hundred meters below the point of the first 
interaction. This effect will slow down the development of the electromagnetic
cascade and shift its maximum.       

In general the shift is limited to no more than 1-2 mean free
paths for the inelastic interaction of the fragments. However, there
might be an additional effect which is, it must be admitted,
speculative. It is based
on the so-called ``{\em sling}'' effect \cite{drem,EW99}. 
and is connected with the deformation of the
shape of the nuclear fragment, which rotates with a high spin 
\cite{loiu}.
As a result of such deformation one can expect an increase in the
mean free path for inelastic interaction which enables the
excited nuclear fragments to penetrate deeper into the atmosphere, slowing
down the cascade development and resulting in a further shift of
$X_{max}$. There will be an additional increase of $N_e$ for nucleus-induced
cascades like that shown in Figure 2b by the dotted line.

All the consequences of the 'sling' effect: polarisation of the
secondary nuclear fragments, deformed shape
of the nuclei rotating with a high spin and cross-section
fluctuations of the excited nuclei are true effects observed at low
energies \cite{satc}. The problem is whether they still hold at
higher energies. However, even if they don't, the de-exitation of the 
nuclear fragments itself will give the needed shift of $X_{max}$ 
if their lifetime is not less than $10^{-12}$ sec.

It is difficult to make reliable numerical estimates of the shift
since there is not enough experimental data and no good theory of
the effect. 
{\em We think that to check our assumption the study of the properties of the
secondary nuclear fragments should be made in high energy
AA-collisions with different nuclear beams.}

{\Large {\bf 6. Discussion}}

{\large {\bf 6.1. Application of the new model to PeV energies}}

{\bf 6.1.1. The mass composition problem}

In order to get better consistency of the results on the primary
cosmic ray mass composition, the combination of two effects is
needed. The increase of only $K^{AA}_{\gamma}$ is not enough because
although it results in a decrease of $N_{\mu}$ and $N_h$ it also leads
to a fall in 
$N_e$ if the measurements are made at sea level, whereas we need an
increase in $N_e$. The slowing down of the electromagnetic cascade alone
does not change the ratio between muons and hadrons. The slowing down
of just the nuclear and electromagnetic cascades (~{\em e.g.} by increasing
the mean free path for nucleon interaction $\lambda_N$ without
increasing $K^N_\gamma$~) results in an increase in both $N_e$ and $N_h$.
Only the combination of both effects: the increase of $K^N_\gamma$ and
the slowing down of the development of the electromagnetic cascade 
changes the energy balance in the 
needed direction: $N_{\mu}$ and $N_h$ decreased, $N_e$ and $X_{max}$ 
increased (~Fig.3~). The required changes of $K^{AA}_{\gamma}$ and 
$X_{max}$ are small, because the sensitivity of the energy balance to
these changes is rather high. 

The result should be a coming together of the $\langle lnA \rangle$
values in Figures 1a and 1b. There will be an increase in the values
in Figure 1a and a small reduction in those in Figure 1b. In view of
the problems associated with $N_\mu^\ast$ (~see \S3~) most of the
predictions in Figure 1b should be disregarded; the 'best estimate' is
then just a little below the crosses in Figure 1b - these values are
reassuringly close to our latest estimate indicated by the upper dashed curve
in Figure 1.

{\bf 6.1.2. The $X_{max}$ controversy}

Although the $X_{max}$ aspect for EAS is one of some complexity we
discuss it briefly. The assumption of a slowing down of the cascade 
development and the consequent shift of $X_{max}$ in $AA$-interactions 
helps us to understand the striking difference 
between the mass composition derived recently by Swordy and Kieda
\cite{swor} and Fowler et al. \cite{fowl}, using Cherenkov light
measurements and those by Roth et al. \cite{roth} and others,
without. Those with Cherenkov light
give a significantly lighter mass composition and lower intensity of the 
primary energy spectrum than those obtained without the use of $X_{max}$ 
\cite{hoer}. If there is indeed a shift
of $X_{max}$ and a consequent increase of $N_e$ the shower looks
more like that initiated by proton and its primary energy, based on the
on-ground measurements is underestimated. Insofar as the $N_e$ values
are most in doubt, from the prediction standpoint the real mass composition
should be closer to that derived by using direct measurements in
the stratosphere or using EAS muons and hadrons on the ground (~Fig.1b~).

{\large {\bf 6.2. Application of the new model to higher energies}}

If the increase of $K^{AA}_{\gamma}$ and the slowing down of the
development of the cascade are due to electron-positron pair
production, and de-excitation of nuclear fragments, there will be 
significant further consequencies of the model because of its charge 
and energy dependent
QED effect and the displacement of the depth of shower maximum. Both
these effects will grow with increasing collision energy. In this case 
it is not
surprising that the inconsistency in the estimates of the primary mass 
composition grows with energy, as noticed by Roth et
al. \cite{roth} and indicated in Figure 1c. 

Of greatest importance is the mass composition at the highest
energies, where, due to the low anisotropy and the rapid growth of
$X_{max}$ with energy, particles are generally considered to be
extragalactic (EG). There is controversy, here, between those who subscribe
to the common view that the EG particles are protons 
(~{\em eg} \cite{bird,abuz,ave} and ourselves (~{\em eg} \cite{WW},
following much earlier work by Tkaczyk et al. \cite{tkac}). A later
publication will deal with this aspect in detail but a few brief
remarks here are in order. 

(i) The change of $K_\gamma^{AA}$ will be even bigger at the highest
energies (~say above 10$^{10}$ GeV~) and the result will be an increase
in the mean primary mass: this follows from the analysis of Ave
et.al.\cite{ave}, in which inclined muon showers were detected - the
corresponding primary energy will now be higher and the rates more in
accord with expectation for a significant fraction of primary iron
nuclei, a recent recalibration of Haverah Park energies \cite{ave}
leads to a reduction in primary energy and an even further
increase in the iron fraction.

(ii) The displacement of the depth of maximum to greater values will,
again, lead to an increase in the mean primary mass as indicated by
the $X_{max}$ values from the experimental data (~see Fig.1b~).

{\Large {\bf 7. Conclusions}}

Our analysis of the present situation with the high energy
interaction models indicates that there is no support for the
introduction of 
a radical change of the models, sometimes proposed
to explain the origin of the knee in the primary cosmic ray energy
spectrum, However, the inconsistencies in the
interpretation of the experimental data on the primary mass
composition, obtained when different
EAS components are used for the analysis, indicate the need for some
improvement to the models which were used hitherto. We propose that
the most promising way is to introduce an additional ( a few percent )
energy transfer into the EAS electromagnetic component combined with a 
slowing down of the cascade development at its initial
stages, which is followed by a small (~$\sim 20-30 gcm^{-2}$~) shift
of $X_{max}$ into the deeper atmosphere and the consequent increase
of $N_e$. The most likely processes
which can be responsible for such changes (~e.g. electron-positron
pair production including multiple pairs, direct photons from the 
hypothetical quark-gluon plasma,
excitation of the secondary nuclear fragments~) are those which occur in
nucleus-nucleus collisions and they should indeed be present {\em at
some level}. The importance of these processes is
expected to grow with energy and offer the hope of resolving some
controversies at very high energies. 

{\Large {\bf Acknowledgements}}

The authors are greatful to the UK's Particle Physics and Astronomy
Research Council and to The Royal Society for financial support. Also we
thank Professors G.Schatz, H.Rebel, K.H.Kampert, J.Kempa and R.Chapman
for useful discussions.


\begin{thebibliography}{99}
\bibitem{heck} Heck D. et al., Proc. 27th Int. Cosm. Ray Conf., 
Hamburg, {\bf1} (~2001~) 233 
\bibitem{sciu} Sciutto S.J., Proc. 27th Int. Cosm. Ray Conf.,
Hamburg, {\bf 1} (~2001~) 237 
\bibitem{EW98} Erlykin A.D., Wolfendale A.W., Astropart. Phys.,
{\bf 9} (~1998~) 213 
\bibitem{anto99} Antoni T. et al. J. Phys.G: Nucl., Part. Phys., {\bf
25} (~1999~) 2161  
\bibitem{swor} Swordy S.P., Kieda D.B., Astropart. Phys., {\bf 13}
(~2000~) 137 
\bibitem{fowl} Fowler J.W. et al., Astropart. Phys., {\bf 15} (~2001~) 49 
\bibitem{roth} Roth M. et al., Proc. 27th Int. Cosm. Ray Conf.,
Hamburg, {\bf1} (~2001~) 88; astro-ph/0102443 
\bibitem{anto01} Antoni T. et al., astro-ph/0106494 (~2001~) 
\bibitem{clay} Clay R.W. et al., Publ. Astron. Soc. Aust., {\bf 15}
(~1998~) 208 
\bibitem{ELW98} Erlykin A.D., Lipsky M., Wolfendale A.W.,
Astropart. Phys., {\bf8} (~1998~) 283 
\bibitem{EW01} Erlykin A.D., Wolfendale A.W. Adv. Space Res.,
{\bf 27} (~2001~) 803 
\bibitem{bere} Berezinsky V.S., Nucl. Phys. {\bf B380} (~2002~) 478 
\bibitem{niko01} Nikolsky S.I. Proc. 27th Int. Cosm. Ray Conf.,
Hamburg, {\bf 4} (~2001~) 1389 
\bibitem{niko95} Nikolsky S.I. Nucl. Phys. B ( Proc. Suppl.),
{\bf 39A} (~1995~) 228  
\bibitem{stam} Stamenov J.N. et al., Proc. 18th Int. Cosm. Ray Conf.,
Bangalore, {\bf 2} (~1983~) 111 
\bibitem{roma} Romakhin V.A., private communication (~1995~)
\bibitem{nava} Navarra G., Nucl. Phys.B ( Proc. Suppl.), {\bf 60B}
(~1998~) 105 
\bibitem{agli} Aglietta M. et al., Astropart. Phys., {\bf 10}
(~1999~) 1 
\bibitem{mits} Mitsui K. et al., Astropart. Phys., {\bf 3} (~1995~) 125 
\bibitem{glas} Glasstetter R. et al., Proc. 26th Int. Cosm. Ray
Conf., Salt Lake City, {\bf 1} (~1999~) 222 
\bibitem{haun} Haungs A. et al., Proc. 27th Int. Cosm. Ray
Conf., Hamburg, {\bf 1} (~2001~) 63 
\bibitem{fomi} Fomin Yu.A. et al., Proc. 27th Int. Cosm. Ray
Conf., Hamburg, {\bf 1} (~2001~) 80 
\bibitem{gres} Gress O. et al., Nucl. Phys.B ( Proc. Suppl.),
{\bf 75A} (~1999~) 299 
\bibitem{arqu} Arqueros F. et al., Astron. Astrophys., {\bf 359} (~2000~) 682 
\bibitem{pali} Paling S., PhD Thesis, University of Leeds (~1997~)
( unpublished ) 
\bibitem{knur} Knurenko S. et al., Proc. 27th Int. Cosm. Ray
Conf., Hamburg, {\bf 1} (~2001~) 145 
\bibitem{petr} Petrukhin A.A., Proc. 27th Int. Cosm. Ray
Conf., Hamburg, {\bf 5} (~2001~) 1768 
\bibitem{kaza} Kazanas D., Nicolaidis A., Proc. 27th Int. Cosm. Ray
Conf., Hamburg, {\bf 5} (~2001~) 1760; astro-ph/0103147, hep-ph/0109247 
\bibitem{chi} Chi X. et al., Astropart. Phys., {\bf 1} (~1993~) 239 
\bibitem{wigm} Wigmans R., astro-ph/0107263 (~2001~) 
\bibitem{ehrl} Ehrlich R., Phys. Rev. D, {\bf 60} (~1999~) 73005 
\bibitem{knap} Knapp J., Heck D., Schatz G. Wissenschaftliche
Berichte FZKA 5828 (~1996~)
\bibitem{riss} Risse M. et al., Proc. 26th Int.Cosm.Ray Conf.,
Salt Lake City, {\bf 1} (~1999~) 135 
\bibitem{cata} Catalano O. et al., Proc. 27th Int. Cosm. Ray
Conf., Hamburg, {\bf 2} (~2001~) 498 
\bibitem{dani83} Danilova T.V., Erlykin A.D., Proc. 18th
Int. Cosm. Ray Conf., Bangalore, {\bf 5} (~1983~) 262 
\bibitem{dani93} Danilova T.V. et al., Journ. Phys. G: Nucl.,
Part. Phys., {\bf19} (~1993~) 429 
\bibitem{erly94} Erlykin A.D., Proc. 1st Int. Symp. on Cosm. Ray
Phys. in Tibet, Lhasa, (~1994~) 74 
\bibitem{erly75} Erlykin A.D., Proc. 14th Int. Cosm. Ray Conf., 
M\"unchen, {\bf7} (~1975~) 2173 
\bibitem{baur} Baur G. et al., Journ. Phys. G: Nucl.,
Part. Phys. {\bf 24} (~1998~) 1657; nucl-th/9606011; hep-ph/0112211  
\bibitem{lee} Lee R.N. et al., hep-ph/0108014 (~2001~)
\bibitem{agak} Agakishev G. et al., Nucl. Phys. A, {\bf 661} (~1999~) 23 
\bibitem{fili} Filimonov K. et al., nucl-ex/0109017 (~2001~)
\bibitem{fein} Feinberg E.L., Nuovo Cim. {\bf 34A} (~1975~) 391 
\bibitem{agga} Aggarwal M.M. et al., Phys. Rev. {\bf C56} (~1997~) 1160 
\bibitem{peit} Peitzmann T., Thoma M.H., hep-ph/0111114 (~2001~) 
\bibitem{cape} Capella A., Nucl. Phys. B, {\bf 60} (~1998~) 138 
\bibitem{capd} Capdevielle J.N. et al., Proc. 27th
Int. Cosm. Ray Conf., {\bf1} (~2001~) 319 
\bibitem{gior} Giorgini M., Manzoor S., hep-ex/0104019 (~2001~)
\bibitem{rama} Ramaty R. et al., Astrophys. J. Suppl., {\bf 40} (~1979~) 487 
\bibitem{drem} Dremin I.M., Man'ko V.I., N.Cim,, {\bf 111A} (~1998~) 439 
\bibitem{EW99} Erlykin A.D., Wolfendale A.W., Nucl. Phys. B 
( Proc. Suppl.), {\bf 75A} (~1999~) 209 
\bibitem{loiu} Lo Iudice N., La Rivista del Nuovo Cimento, {\bf 9} (~2000~) 1 
\bibitem{satc} Satchler G.R., Introduction to Nuclear Reactions,
Macm. Press Ltd., NY (~1980~)
\bibitem{hoer} H\"orandel J.R., Proc. 27th Int. Cosm. Ray Conf., 
Hamburg, {\bf1} (~2001~) 71 
\bibitem{bird} Bird D.J. et al., Phys. Rev. Lett., {\bf 71} (~1993~) 3401 
\bibitem{abuz} Abu-Zayad T. et al., astro-ph/9911144 (~1999~)
\bibitem{ave} Ave M. et al., Proc. 27th Int. Cosm. Ray Conf.,
Hamburg, {\bf1} (~2001~) 381 and 390 
\bibitem{WW} Wibig T., Wolfendale A.W., Proc. 27th Int. Cosm. Ray Conf.,
Hamburg, {\bf 5} (~2001~) 1987 
\bibitem{tkac} Tkaczyk W. et al., Journ. Phys. A, {\bf 8} (~1975~) 1518 
\end{thebibliography}
\end{document}